\documentclass[useAMS,usenatbib,usegraphicx]{mn2e}
\usepackage{amssymb}
\usepackage{latexsym}
\usepackage {keyval}
\usepackage{amsmath}

\title[H$_2$O maser outflow in the LkH$\alpha$ 234 region]{A very young, compact bipolar H$_2$O maser outflow in the intermediate-mass star-forming LkH$\alpha$~234 region}
\author[Torrelles et al.]{J. M. Torrelles,$^1$\thanks{E-mail: torrelles@ieec.cat}  S. Curiel,$^{2}$ R. Estalella,$^{3}$ G. Anglada,$^4$ J. F. G\'omez,$^4$ J. Cant\'o,$^2$
\newauthor N. A. Patel,$^5$ M. A. Trinidad,$^6$
J. M. Girart,$^7$ C. Carrasco-Gonz\'alez,$^{8}$
\newauthor L. F. Rodr\'{\i}guez$^8$\\
$^{1}$Institut de Ci\`encies de l'Espai (CSIC-IEEC) and Institut de Ci\`encies del Cosmos (UB-IEEC),\\ ~Mart\'{\i} i Franqu\`{e}s 1, 08028 Barcelona, Spain\\
$^{2}$Instituto de Astronom\'{\i}a (UNAM), Apartado 70-264, 04510 M\'exico
D. F., M\'exico\\
$^{3}$Departament d'Astronomia i Meteorologia and Institut de Ci\`{e}ncies del Cosmos (IEEC-UB), Universitat de Barcelona,\\~~Mart\'{\i} i Franqu\`{e}s 1, 08028 Barcelona, Spain\\
$^{4}$Instituto de Astrof\'{\i}sica de Andaluc\'{\i}a (CSIC), Apartado 3004, 18080 Granada, Spain\\
$^{5}$Harvard-Smithsonian Center for Astrophysics, 60 Garden Street, Cambridge, MA 02138, USA\\
$^{6}$Departamento de Astronom\'{\i}a, Universidad de Guanajuato, Apdo. Postal 144, 36000 Guanajuato, M\'exico\\
$^{7}$Institut de Ci\`encies de l'Espai (CSIC-IEEC), Campus UAB, Facultat de Ci\`encies, C5p 2, 08193 Bellaterra, Spain\\
$^{8}$Centro de Radioastronom\'{\i}a y Astrof\'{\i}sica (UNAM), 58089 Morelia, M\'exico}
\begin{document}

\date{Accepted 2014 April 23. Received 2014 April 21; in original form 2014 April 1}

\pagerange{\pageref{firstpage}--\pageref{lastpage}} \pubyear{2014}

\maketitle

\label{firstpage}

\begin{abstract}

We report multi-epoch VLBI H$_2$O maser observations towards the compact cluster of YSOs close to the Herbig Be star LkH$\alpha$~234. This cluster includes LkH$\alpha$~234 and at least nine more YSOs that are formed within projected distances of $\sim$10~arcsec ($\sim$9,000~au). We detect H$_2$O maser emission towards four of these YSOs. In particular, our VLBI observations (including proper motion measurements) reveal a remarkable very compact ($\sim$0.2~arcsec = $\sim$180~au), bipolar H$_2$O maser outflow emerging from the embedded YSO VLA~2. We estimate a kinematic age of  $\sim$40~yr for this bipolar outflow, with expanding velocities of $\sim$20~km~s$^{-1}$
and momentum rate $\dot M_w V_w$ $\simeq$ $10^{-4}-10^{-3}$ M$_{\odot}$~yr$^{-1}$~km~s$^{-1}$$\times (\Omega$/$4\pi)$, powered by a YSO of a few solar masses.
We propose that the outflow is produced by recurrent episodic jet ejections associated with the formation of this YSO. Short-lived episodic ejection events have previously been found towards high-mass YSOs. We show now that this behaviour is also present in intermediate-mass YSOs. These short-lived episodic ejections are probably related to episodic increases in the accretion rate, as observed in low-mass YSOs. We predict the presence of an accretion disk associated with VLA~2. If detected, this would represent one of the few known examples of intermediate-mass stars with a disk-YSO-jet system at scales of a few hundred au.

\end{abstract}


\section{Introduction}

The early stages of evolution of low-mass stars are relatively well characterised by the formation of a system with a central protostar, accreting material from a rotating accretion disk at scales of a few hundreds of astronomical units (au), and simultaneously ejecting a collimated outflow with the presence and crucial role of magnetic fields. Accretion and mass-loss processes, two closely related mechanisms,  govern the formation of low-mass stars (e.g., Girart, Rao \& Marrone 2006; McKee \& Ostriker 2007; Machida, Inutsuka \& Matsumoto 2008; Armitage 2011; Williams \& Cieza 2011). In fact, spatio-kinematical studies of jets and Herbig-Haro (HH) objects show that outflows from low-mass young stellar objects (YSOs) are non-steady, but presenting variability in the ejection velocity as well as pulsed events, probably related to recurrent instabilities in the accretion disks (e.g., Zinnecker, McCaughrean \& Rayner 1998;  Reipurth \& Bally 2001; Estalella et al. 2012). With respect to the formation of massive stars ($\gtrsim$ 10~M$_{\odot}$), there are several examples showing the presence of massive disk-protostar-jet systems at scales of thousand au (in some cases with magnetic fields oriented parallel to the collimated outflows), indicating that stars at least up to $\sim$20~M$_{\odot}$ form via an accretion disk as low-mass do (e.g., Patel et al. 2005; Jim\'enez-Serra
et al. 2007; Torrelles et al. 2007; Davies et
al. 2010; Carrasco-Gonz\'alez et al. 2010, 2012a; Vlemmings et al. 2010; Fern\'andez-L\'opez
et al. 2011; Maud \& Hoare 2013). Furthermore, variation in the ejection velocity of outflows as well as remarkable short-lived episodic ejection events (few tens of years) have been reported towards massive YSOs through radio continuum
(e.g., Mart\'{\i} et al 1995; Curiel et al. 2006) and H$_2$O maser spatio-kinematical observations (e.g., Torrelles et al. 2001,  2003; Surcis et al. 2011a,b; Chibueze et al. 2012; Sanna et al.  2012;  Kim et al. 2013; Trinidad et al.  2013).  This indicates that outflows in massive YSOs are also non-steady, probably due to instabilities in the accretion processes as in the case of low-mass YSOs.

Mainly because most of the observational efforts have been concentrated in low- and high-mass YSOs, little is known on what happens in intermediate-mass protostars, with only a few examples showing clear disk-YSO-jet systems at scales of a few hundred au (e.g., Palau et al. 2011; Carrasco-Gonz\'alez et al. 2012b), but up to now without any observational information on the magnetic field and outflow kinematical behaviour at these small scales. 
Observations of the first stages of the evolution of intermediate-mass YSOs at small scales are therefore important to fill that gap, and thus to have a more complete vision about the processes of star formation throughout a more continuous range of masses. In this paper, we present multi-epoch Very Long Baseline Interferometry (VLBI) H$_2$O maser observations towards intermediate-mass YSOs to study the spatio-kinematical distribution of their masers with $\sim$0.5~mas resolution. This kind of study has previously been very useful in massive star-forming regions to detect outflow activity through the presence of H$_2$O maser emission, allowing the identification of new centres of star formation, some of them associated, as mentioned before, with short-lived episodic ejection events with kinematic ages of a few tens of years. With the present work, we now extend this study to intermediate-mass star-forming regions.

The optically visible Herbig Be star LkH$\alpha$~234, with luminosity  $\sim$10$^3$~L$_{\odot}$ and mass $\sim$5~M$_{\odot}$, is one of the brightest sources of the NGC~7129 star-forming region, located at a distance of 0.9~kpc (Herbig 1960; Strom et al. 1972; Bechis et al. 1978; Fuente et al. 2001; Trinidad et al. 2004; Xu et al. 2013). High angular resolution near- and mid-infrared images obtained by Kato et al. (2011) reveal, in addition to LkH$\alpha$~234, a cluster of eight YSO candidates within a distance of $\sim$10 arcsec from LkH$\alpha$~234 (named by Kato et al. as objects B, C, D, E, F, G, NW1, and NW2). Very Large Array (VLA) observations show five radio continuum sources in a region of $\sim$5 arcsec (VLA~1, VLA~2, VLA~3A, VLA~3B, LkH$\alpha$~234; Trinidad et al. 2004), some of them with mid-infrared counterparts: VLA~2 (NW2), VLA~3A+3B (NW1), and LkH$\alpha$~234 (see Fig. 1). NW2 and NW1, are not seen in the J, H, and K-band images, but they are bright in the $L'$,  $M'$, and mid-infrared bands, indicating that they are highly embedded YSOs (Kato et al. 2011). 
H$_2$O maser emission is detected towards this region, mainly associated with VLA~1, VLA~2 (NW2), and VLA~3A+3B (NW1) (Tofani et al. 1995; Umemoto et al. 2002; Trinidad et al. 2004; Marvel 2005; see Fig. 1). Several outflows emerging from this region have been also observed, e.g.: i) a jet observed in [S II], extending to the south-west (p.a. $\sim$  252$^{\circ}$) up to distances of $\sim$40 arcsec ($\sim$36,000~au), without evidence of a counter-jet  (Ray et al. 1990);  ii) an infrared jet seen in v=1--0 S(1) H$_2$ emission, extending towards the south-west (p.a. $\sim$ 226$^{\circ}$) up to distances of $\sim$10 arcsec ($\sim$9,000~au), also without any evidence of a counter-jet (Cabrit et al. 1997). The overlapping  of these two outflows within a small area of the sky makes it very difficult to clearly distinguish between various candidate driving sources. What we know, however, is that all the different studies exclude that the Herbig Be star LkH$\alpha$~234 is driving any of the outflows of this star-forming region, but that the main outflow activity centres are probably the YSOs VLA~2 (NW2) and VLA~3A+3B (NW1) (Trinidad et al. 2004; Marvel 2005; Kato et al. 2011). Following the nomenclature of Trinidad et al. (2004), in this paper we will refer to the region containing all this cluster of YSOs as the ``star-forming LkH$\alpha$~234 region".

\begin{figure*}
 \centering
 \includegraphics[width=140mm, clip=true]{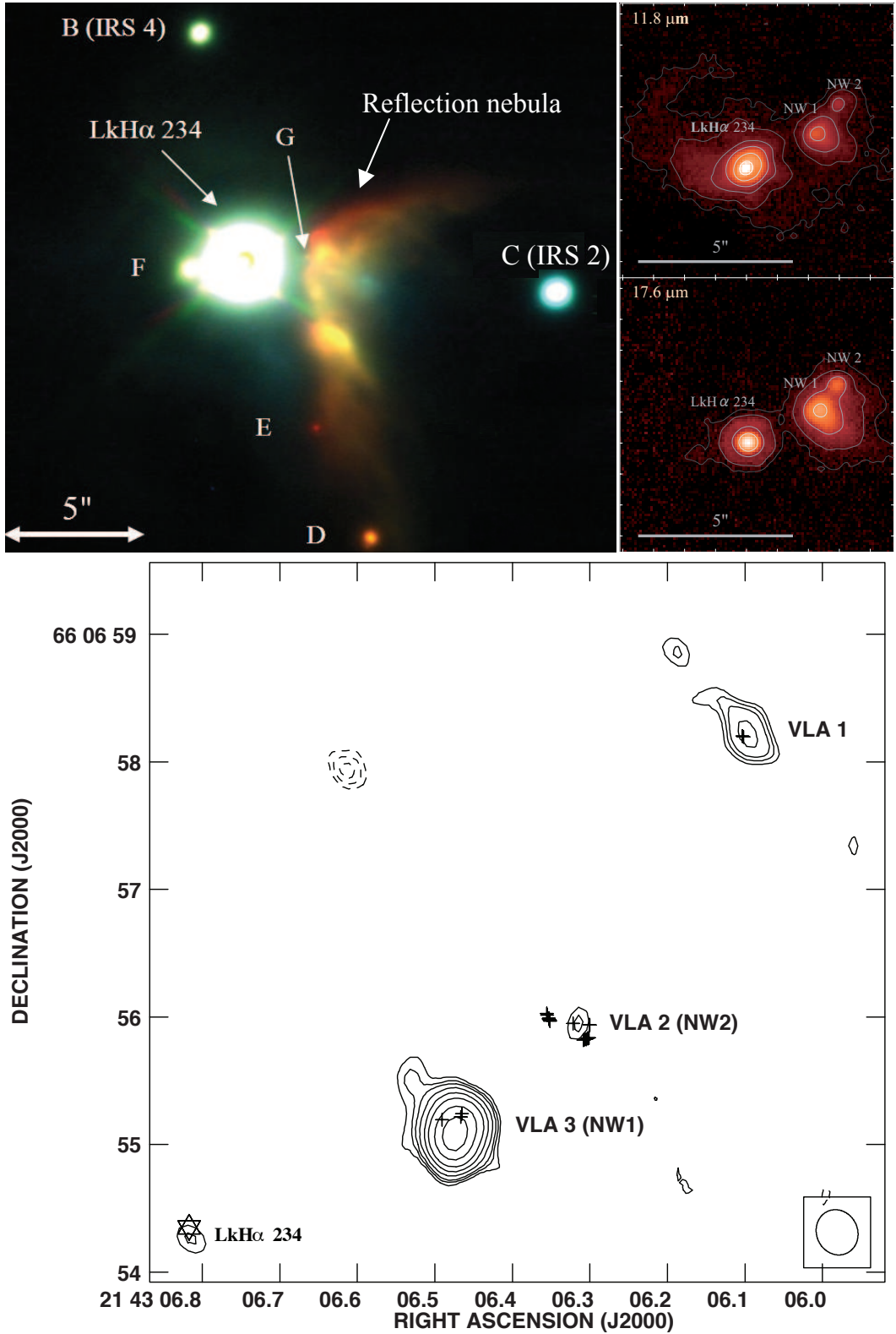} 
 \caption{JHK colour composite image ({\it top left-hand panel}), and  11.8 $\mu$m and 17.6~$\mu$m Keck LWS images ({\it top right-hand panels}) around the Herbig Be star LkH$\alpha$~234. In addition to LkH$\alpha$~234, eight YSO candidates are also indicated (B, C, D, E, F, G, NW1, NW2; images and nomenclature from Kato et al. 2011). {\it Bottom panel}: VLA-3.6~cm continuum contour map of the region showing emission from four sources, VLA~1, VLA~2, VLA~3 (this source, when observed with higher angular resolution, has two components, VLA~3A and 3B, which are separated by $\sim$0.3~arcsec; Trinidad et al. 2004), and LkH$\alpha$~234 (the optical position is indicated by a star). The crosses indicate the position of the H$_2$O masers detected with the VLA (epoch 1999 June 29; observations and map from Trinidad et al. 2004). The mid-infrared sources NW1 and NW2 detected by Kato et al. (2011) coincide in position with the radio continuum sources VLA~3 (A+B) and VLA~2, respectively.}
 \end{figure*}

In Section~2, we present the multi-epoch VLBI H$_2$O  maser observations towards  the star-forming LkH$\alpha$~234 region and the main results, in particular the discovery of a very young, compact bipolar H$_2$O maser outflow associated with VLA~2 (NW2). The implications of these results are discussed in Section~3, while the main conclusions of this work are presented in Section~4.

\section{Observations and results}

 The $6_{16}-5_{23}$ H$_2$O maser transition (rest 
frequency = 22235.08~MHz) was observed with the Very Long Baseline Array
(VLBA) of the National Radio Astronomy Observatory (NRAO)\footnote{The NRAO is a facility
of the National Science Foundation operated under  cooperative agreement by Associated
Universities, Inc.} towards the star-forming LkH$\alpha$~234 region at three epochs (2001 December 2, 2002 February 11,
and 2002 March 5), during an observing on-source time of $\sim$5~hours per epoch. A bandwidth of 8~MHz sampled over 512 spectral line channels of $\sim$0.21~km s$^{-1}$ width and centred at $V_{\mathrm LSR} = -9.3$~km~s$^{-1}$ was used. The data were correlated at the NRAO Pete V. Domenici Science Operations Center (SOC). Calibration and imaging was made using the Astronomical Image Processing System (AIPS) package of NRAO. Delay and phase calibration was provided by the sources 3C345, 3C454.3, BL Lac, B1739+522, and B2007+777, while bandpass corrections were determined through observations of 3C345, 3C454.3, and BL Lac. The size of the synthesised beam was $\sim$0.4~mas for the three epochs. Self-calibration of the uv data was made through a strong point-like maser persisting in all three epochs. This maser, with flux density $\sim$10~Jy and radial velocity $V_\mathrm{LSR}= -10.1$~km s$^{-1}$, has absolute coordinates $\alpha$(J2000.0) = {\rm 21$^h$43$^m$06.312$^s$}, $\delta$(J2000.0)
= 66$^{\circ}$06$'$55.79$''$ ($\pm$ 0.05$''$), and is spatially associated with the VLA~2 source. 
This maser was also used for a first preliminary coordinate alignment of the three epochs of observations, which helped us to the identification of the H$_2$O maser emission in the region.

We detected H$_2$O maser emission towards the sources VLA~1, VLA~2, and VLA~3A+3B (Fig. 2). 
For each epoch, we obtained individual data cubes of 8096$\times$8096 pixels $\times$ 512 velocity channels, and pixel size of 0.1~mas, for three fields, centred on each of the radio continuum sources. 
The rms noise level of the individual velocity channel maps depends on the intensity of the masers in each velocity channel, and ranges from $\sim$7 to 25~mJy~beam$^{-1}$. For the identification of the maser spots in the region (by maser spot we mean emission occurring at a given velocity channel and distinct spatial position), we adopted a signal-to-noise ratio (SNR) threshold of $\gtrsim$ 10. All the maser spots detected in the region, most of them unresolved with the beam size of $\sim$0.4~mas, were fitted with two-dimensional Gaussian profiles, obtaining their positions, flux density, and radial velocity.
From the SNR  and the beam size, we estimate that the 1$\sigma$ accuracy in the relative positions of the maser spots within each epoch is $\lesssim$ 0.02~mas (beam/[2$\times$SNR]; Meehan et al. 1998).

\begin{figure}
 \centering
 \includegraphics[width=83mm, clip=true]{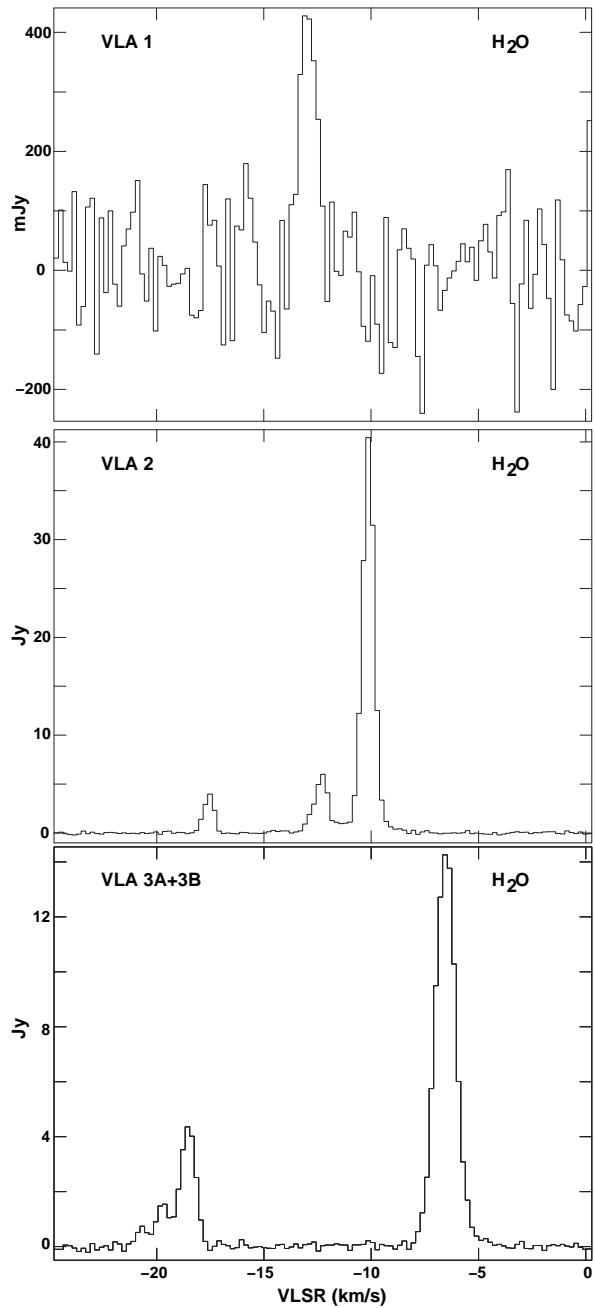} 
 \caption{H$_2$O maser spectra observed with the VLBA towards VLA~1 (top), VLA~2 (middle), and VLA~3A+3B (bottom) in 2002 February 11. These spectra were obtained by adding components from the clean images of the different regions with the task ISPEC (AIPS). For VLA~1, H$_2$O maser emission is only detected in this particular epoch. For VLA~3A+3B, weak emission ($\sim$0.09~Jy) is also detected at $V_\mathrm{LSR}\simeq -37$~km~s$^{-1}$ in epoch 2001 December 2.}
 \end{figure}
 
 \begin{table*}
 \centering
 \begin{minipage}{140mm}
    \caption{Proper motions of the H$_2$O maser features towards VLA~2 and VLA~3 in the LkH$\alpha$~234 region$^a$}
 \begin{tabular}{@{}lrrcrrcc@{}}
  \hline
Feature &Detected& V$_{LSR}$ & Flux density & $\Delta x$& $\Delta y$ & $V_x$ & $V_y$ \\
ID &epochs& (km~s$^{-1}$) & (Jy) & (mas) & (mas) & (km~s$^{-1}$) & (km~s$^{-1}$)\\
    \hline
   VLA~2 & & & & & & &\\ 
   \hline
1&1,2,3  &  $-$12.5  & 0.36 & $-$159.19 &   $-$84.48 &  $-$29.2 $\pm$ 0.1   & $-$12.2 $\pm$ 0.4\\
2&1,2,3  &   $-$17.7    &   0.40 &  $-$134.35 &   $-$20.55 & $-$14.0  $\pm$  0.1  &  3.9   $\pm$ 0.3\\
3$^b$&1,2,3  &   $-$17.5    &   2.36 &   $-$133.30 &   $-$10.51 & $-$12.9  $\pm$  0.1  &  $-$2.1   $\pm$  0.3\\
4$^b$&1,2,3  &   $-$18.2    &   0.09 &  $-$132.95 &   $-$10.95 & $-$13.9  $\pm$  0.2  & $-$3.2   $\pm$  0.6\\
5 &2,3   &   $-$9.5     &  0.17  & $-$130.09  & $-$107.94  & $-$21.6   & $-$21.8  \\
6&1,2,3  &   $-$11.6    &   0.16 &     $-$129.00 &   $-$116.10 &  $-$9.4  $\pm$ 0.1   & $-$7.1   $\pm$  0.1\\
7 &2,3   &   $-$9.9     &  0.14  & $-$113.12  & $-$130.15  & $-$20.5  & $-$2.0  \\
 8&2,3   &   $-$9.3     &  0.12  & $-$108.56  & $-$132.86  & $-$5.7  & $-$4.4 \\
9&1,2,3  &   $-$10.6    &   0.26 &  $-$105.51 &  $-$130.59 &   $-$3.4  $\pm$    0.9 & $-$5.5   $\pm$ 0.4 \\
 10&2,3   &  $-$10.8     &  0.16  & $-$104.42  & $-$130.83  & $-$1.4  & $-$3.6 \\
 11&2,3   &  $-$10.6     &  0.09  & $-$102.04  & $-$130.15  & $-$4.9  & $-$6.0 \\
 12&2,3   &   $-$8.7     &  0.14  &  $-$85.68  & $-$107.66  & $-$13.7   & $-$7.1 \\
 13&1,2,3  &   $-$13.5    &   0.09 &   $-$84.49 &    $-$77.90 & $-$10.8  $\pm$  1.3   & $-$10.4  $\pm$ 1.1\\
 14&1,2,3  &    $-$9.9    &   1.84 &   $-$69.59 &  $-$135.08 &  $-$6.3  $\pm$  0.2  & $-$7.0   $\pm$  0.3\\
15 &1,2,3  &    $-$9.9    &   3.84 &   $-$67.62 &  $-$134.82 &  $-$8.3  $\pm$ 0.1  &  $-$8.1   $\pm$ 0.1 \\
 16&1,2,3  &    $-$9.9    &   6.58 &   $-$66.84 &  $-$134.67 &  $-$1.2  $\pm$  0.1  & $-$6.2   $\pm$  0.1\\
 17&1,2,3  &   $-$10.6    &   0.24 &   $-$65.22 &  $-$134.56 &  $-$3.7 $\pm$  0.5  & $-$5.8   $\pm$ 0.1\\
 18&1,2,3  &   $-$10.6    &   0.40 &   $-$64.49 &  $-$134.28 &   1.7  $\pm$  0.3  & $-$6.9   $\pm$  0.3\\
 19&1,2,3  &   $-$10.6    &   0.28 &   $-$63.95 &   $-$134.20 &   5.0  $\pm$  0.3  & $-$5.8   $\pm$  0.1\\
  20& 2,3   &  $-$11.2     &  0.08  &  $-$49.28  & $-$128.88  &  $-$1.3   & $-$4.2  \\
 21&1,2,3  &   $-$12.9    &   0.51 &   183.46 &    119.90 &  16.9  $\pm$  1.3   & 11.9   $\pm$ 0.1\\
 22&1,2,3  &    $-$8.9    &   0.19 &    200.10 &    123.60 &  15.0  $\pm$  0.8  &       0.0   $\pm$  1.5\\
 23 &2,3   &  $-$9.5     &  0.11  &  207.28  &   94.83  & 24.2    & $-$1.1 \\
 24 &2,3   &   $-$9.7     &  0.14 &  207.45  &   94.04  & 21.3     & 19.4\\
  25&1,2,3  &    $-$9.3    &   0.12 &   207.52 &     93.10 &  20.4  $\pm$   0.4  & 15.3   $\pm$ 1.1\\
  26&1,2,3  &   $-$14.8    &   0.16 &    207.70 &    77.21 &  26.8  $\pm$  0.3  & 13.4   $\pm$  0.6\\
  27&1,2,3  &   $-$13.3    &   0.18 &   208.51 &    74.28 &  24.8  $\pm$  0.1  &  7.3   $\pm$ 0.1\\
  28&1,2,3  &     $-$11.0    &   0.11 &   208.65 &    77.64 &  31.5  $\pm$  3.8 &   6.4   $\pm$  2.0\\
  29&1,2,3  &   $-$12.5    &   0.24 &   208.68 &    76.81 &  23.7  $\pm$  0.3  & 13.5   $\pm$  0.8\\
   30& 2,3   &  $-$12.7     &  0.30  &  209.11  &   78.74  & 28.0    & 13.6\\
  31&1,2,3  &   $-$12.7    &   0.18 &   209.18 &    74.79 &  24.5  $\pm$   0.2&  7.1   $\pm$ 0.2\\
   32& 2,3   &  $-$11.6     &  0.26  &  210.33  &   77.37  & 24.6  & 15.2\\
  \hline
  VLA~ 3 (A/B) & & & & & & &\\  
   \hline
    33&1,2,3  &   $-$20.9    &   0.11 &   821.72 &  $-$856.41 &    6.0  $\pm$  0.4  &  $-$6.5 $\pm$ 0.1\\
   34& 2,3   &  $-$20.5     &  0.18  &  822.91  & $-$856.48  & $-$8.6  & $-$2.6 \\
  35&1,2,3  &   $-$20.1    &   0.16 &   823.82 &  $-$856.07 &  $-$3.7  $\pm$  0.2  & $-$10.5  $\pm$ 0.5\\
  36&1,2,3  &   $-$19.8    &   0.98&   825.18 &  $-$856.18 &  $-$0.6  $\pm$ 0.5   &  $-$9.0   $\pm$  0.2 \\
   37&2,3   &  $-$19.6     &  0.10  &  826.24  & $-$856.85  & $-$3.6 & $-$0.3 \\
   38&1,2,3  &   $-$18.2    &   1.20 &   826.41 &  $-$853.99 &  $-$2.2  $\pm$ 0.1   & $-$6.1   $\pm$ 0.1 \\
   \hline
   VLA~ 3B & & & & & & &\\  
   \hline
   39&1,2~~  &   $-$4.7     &  0.17  &   878.30  & $-$734.44  & $-$3.9& $-$5.8 \\
  40&1,2,3  &    $-$5.7    &   1.33 &   878.59 &  $-$735.32 &  $-$1.2  $\pm$  0.3 &  4.4   $\pm$  0.3\\
 \hline
 VLA~ 3A & & & & & & &\\  
 \hline
  41 &1,2~~  &  $-$36.9     &  0.09  &  945.19  &  $-$916.78 &   9.7 & 39.9\\

        \hline
\end{tabular}

\noindent $^a$For each feature (numbered in column 1), the
epochs of detection are given in column 2 (epoch 1: 2001 December 2; epoch 2: 
2002 February 11; epoch 3: 2002 March 5). Columns 3 and 4 give, respectively, the radial velocity and
flux density of the masers as observed in the first epoch
of detection. The position offsets of the features, at the epoch where they were first detected, are given in columns 5 and 6.
The (0,0) has absolute coordinates 
RA(J2000) = 21$^h$43$^m$06.323$^s$, DEC(J2000)=
66$^{\circ}$06$'$55.92$''$ ($\pm$ 0.05$''$), coinciding with the peak position of the VLA~2 radio continuum source (see text).
The proper motions of the features and their uncertainties 
are reported in
columns 7 and 8.  For those features detected only in two epochs there are not uncertainties estimates for the proper motion values. From the data we have, it has not been possible to associate some of the masers to sources VLA~3A or VLA~3B (those indicated by VLA~3 (A/B)).\\
$^b$The features 3 and 4 are identified, respectively, as features 25 and 26 listed by Marvel (2005), and were used to align his
data set of epoch 2003 February 23 with ours. In addition these features 3 and 4 are identified with the VLA feature 4 listed by Trinidad et al. (2004) (epoch 1999 June 29) and was used to align the VLA data set with ours (see text and Fig. 5).
\end{minipage}
\end{table*}

For the proper motion measurements of the H$_2$O masers, we have identified a set of maser features persisting in different epochs of our observations. By ``maser feature'' we mean a group of spots with positions coinciding within a beam size as well as radial velocities appearing in consecutive velocity channels (typically within $\sim$1~km~s$^{-1}$). We first measured the proper motion of the maser features 
adopting the maser spot used to self-calibrate the uv data as the reference for the preliminary alignment of the three epochs. Those first steps already showed a clear symmetric bipolar outflow distribution of masers around VLA~2 (see Section 2.1.2). 
However, since the initial reference maser spot may have its own proper
motion, resulting in arbitrary offsets in the proper motions for the
whole system, we have defined a more ``stationary'' reference frame
assuming that the mean position of the maser features around VLA~2
persisting in all three observed epochs remains unchanged with time.
All the proper motion values given in this paper have been determined after this realignment by a linear fitting to the position of the maser features of the different epochs as a function of time (Table 1). 

\subsection{Results}

\subsubsection{Spatial distribution of the H$_2$O masers}

\begin{figure*}
 \centering
 \includegraphics[width=150mm, clip=true]{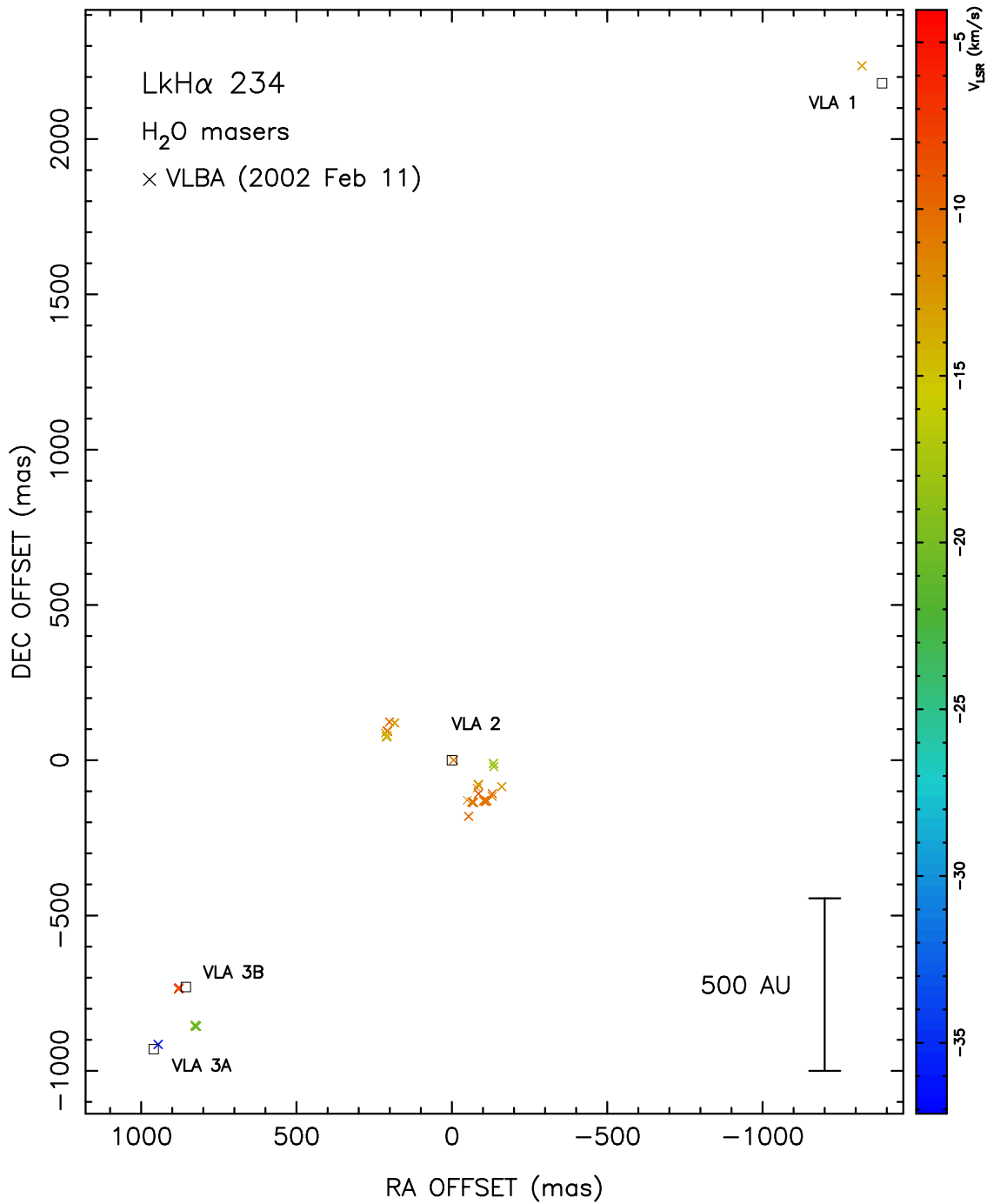} 
 \caption{Positions of the H$_2$O masers measured with the VLBA for epoch 2002 Feb 11. Colour code indicates the radial velocity of the masers in km~s$^{-1}$. The positions of the peak emission of the radio continuum sources VLA~1, VLA~2, VLA~3A, VLA~3B at 3.6~cm (Trinidad et al. 2004) are indicated by open squares. No water maser emission is detected towards the star LkH$\alpha$~234, located at $\sim$ 2~arcsec southeast from VLA~3A (outside of the region represented in this figure). The
 (0,0) has absolute coordinates $\alpha$(J2000) = 21$^h$43$^m$06.323$^s$,  $\delta$(J2000) = 66$^{\circ}$06$'$55.92$''$ ($\pm$ 0.05$''$). The accuracy in the relative positions of the maser spots  is better than $\sim$0.02~mas, while the accuracy in the relative positions of the masers and the radio continuum sources plotted in this figure is $\sim$10-20~mas (see Section 2).}
 
\end{figure*}

H$_2$O maser emission was detected towards the sources VLA~1, VLA~2, and VLA~3A+3B in the three observed epochs, except in VLA~1, where emission was only detected on 2002 February 11. As previously reported by Trinidad et al. (2004) and Marvel (2005), there is no H$_2$O maser emission towards the Herbig Be star LkH$\alpha$~234, consistent with the absence of outflow activity in this object. The maser emission spans a velocity range and flux density from 
$V_\mathrm{LSR}\simeq -13$ 
to $-12$~km~s$^{-1}$, $S_{\nu}$ $\simeq$ 0.07 to 0.3~Jy (VLA~1), $V_\mathrm{LSR}\simeq -18$ 
to $-9$~km~s$^{-1}$, $S_{\nu}$ $\simeq$ 0.07 to 11~Jy  (VLA~2), and $V_\mathrm{LSR}\simeq -37$ 
to $-4$~km~s$^{-1}$,  $S_{\nu}$ $\simeq$ 0.07 to 13~Jy (VLA~3A+3B). In Fig. 2 we show the spectra of the H$_2$O maser emission obtained towards VLA~1, VLA~2, and VLA~3A+3B on 2012 February 11, while the positions and radial velocity distribution of the masers for the same epoch are plotted in Fig. 3 for a general overview of the region. In general, the spatial distribution of the masers that we find with the VLBA agrees with the spatial distribution found by Trinidad et al. (2004) with the VLA in 1999 June 29 but with a beam size of $\sim$ 0.08~arcsec (see Figs. 1 and 3). Specially remarkable is the similarity of the structure found around VLA~2, both with the VLBA (Fig. 3) and the VLA (Fig. 1), with two main groups of masers separated by $\sim$0.4~arcsec ($\sim$360 au) as well as maser emission at the center of these two groups coinciding very well with the centre of the centimetre emission of VLA 2. We do not find a significant segregation in the radial velocity of the northeastern and southwestern groups of masers associated with VLA~2, with the two groups having similar radial velocities.

\begin{figure*}
 \centering
 \includegraphics[width=176mm, clip=true]{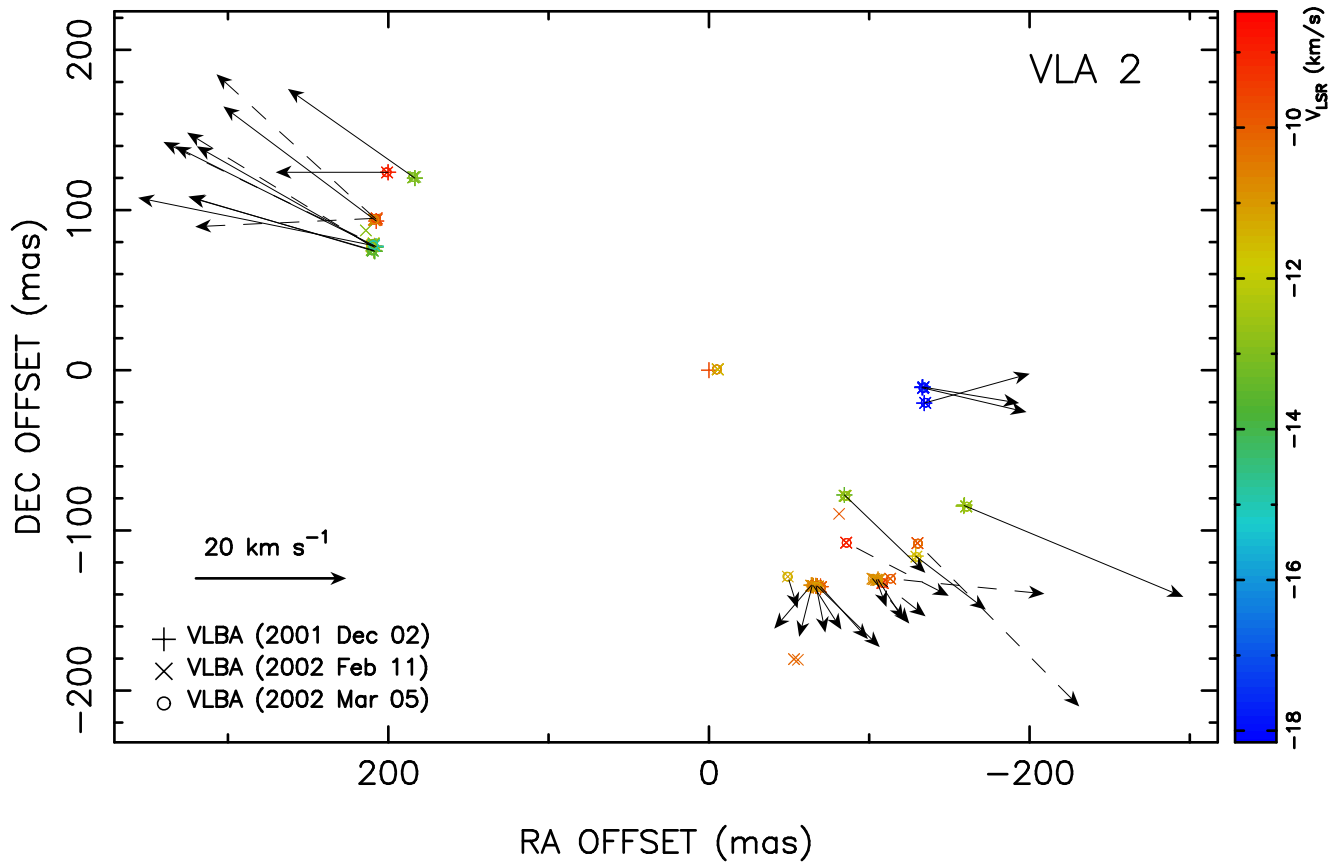} 
 \caption{Positions, radial velocities, and proper motions of the H$_2$O masers associated with VLA~2 as observed with our set of three epochs of VLBA data (2001 December 2, 2002 February 11, and 2002 March 5). The colour scale indicates the radial velocity of the masers.The arrows represent the proper motion vectors of the H$_2$O maser features listed in Table 1. The solid arrows represent proper motions obtained from three epochs, while dashed arrows represent proper motions obtained from only two epochs. The
 (0,0) has absolute coordinates $\alpha$(J2000) = 21$^h$43$^m$06.323$^s$,  $\delta$(J2000) = 66$^{\circ}$06$'$55.92$''$ ($\pm$ 0.05$''$).}
\end{figure*}

\begin{figure*}
 \centering
 \includegraphics[width=176mm, clip=true]{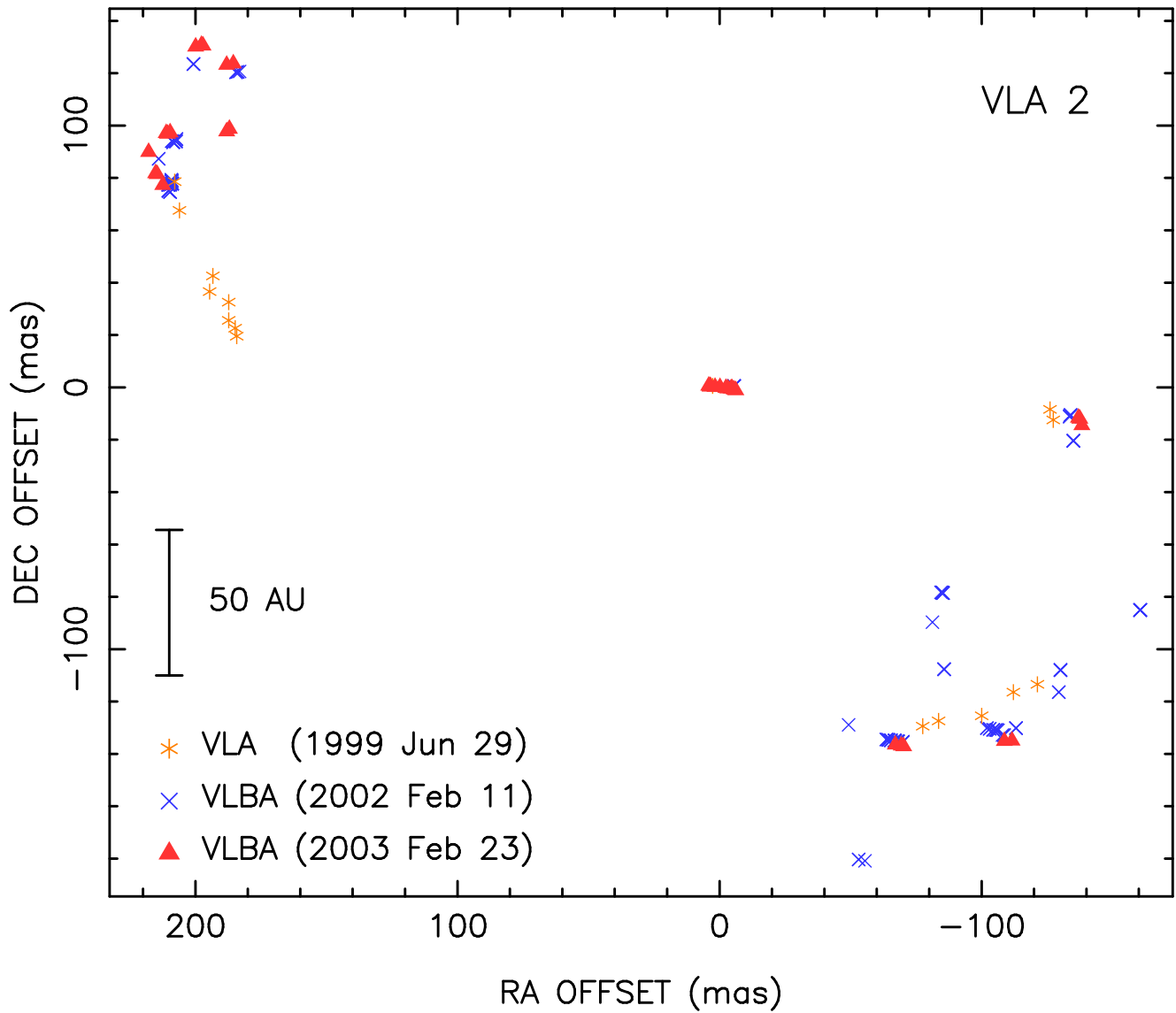} 
 \caption{Positions of the H$_2$O masers associated with VLA~2 as measured with the VLA epoch 1999 June 29 (Trinidad et al. 2004), VLBA  epoch 2002 February 11 (this paper), and VLBA epoch 2003 February 23 (Marvel 2005).  The
 (0,0) has absolute coordinates $\alpha$(J2000) = 21$^h$43$^m$06.323$^s$,  $\delta$(J2000) = 66$^{\circ}$06$'$55.92$''$ ($\pm$ 0.05$''$).}
\end{figure*}

In this paper, all the offset positions of the H$_2$O masers are relative to the maser spot position (0,0) detected with the VLBA at the center of the two groups of masers around VLA~2 in the first epoch of our observations (2001 December 2),  with V$_\mathrm{LSR}$ $\sim$$-$10~km~s$^{-1}$, flux density $\sim$0.2~Jy, and absolute
coordinates RA(J2000) = 21$^h$43$^m$06.323$^s$, DEC(J2000)=
66$^{\circ}$06$'$55.92$''$ ($\pm$ 0.05$''$). We identify this particular maser detected with the VLBA, with the VLA maser reported by Trinidad et al. (2004) at the center of these two groups of masers ($\sim$1~Jy~beam$^{-1}$), with similar velocity ($\sim$$-$11~km~s$^{-1}$) and absolute coordinates (RA[J2000] = 21$^h$43$^m$06.322$^s$, DEC[J2000]=
66$^{\circ}$06$'$55.95$''$),  
and which is located very close to the peak position of the VLA~2 source (within $\sim$10~mas)\footnote{The fact that the VLA~2 source was also detected at 1.3~cm and observed simultaneously with the H$_2$O masers, allowed Trinidad et al. (2004) to estimate the relative positions of the continuum source and the VLA masers with an accuracy of $\sim$10~mas.}. This also allows us to plot in Fig. 3 the relative positions of the different radio continuum sources (VLA~1, VLA~2, VLA~3A, VLA~3B) with respect to the VLBA masers with an accuracy of 10-20 mas, which is high enough for a proper comparison of the maser distribution with respect to the radio continuum sources. In this way, we see from Fig. 3  that, in addition to VLA~1 and VLA~2, both components of VLA~3 (3A and 3B) have associated H$_2$O
maser emission. There is also maser emission in between both binary components, but with the present data we cannot establish whether it is associated to either component
or if it is associated to a different YSO not yet detected.

\subsubsection{Proper motions of the H$_2$O masers}

We measured proper motions of 41 H$_2$O maser features in the region, of which 26 were detected in the three epochs, while 15 were detected in only two epochs. In Table 1 we list the main parameters of these features, including the detected epochs, radial velocity, intensity, offset positions, proper motion values with their uncertainties, and their association to any of the sources in the region. Most of the maser features where we measured proper motions are associated with VLA~2 (32), while nine are associated with VLA~3A+B. Proper motion measurements  of the masers towards VLA~1 were not possible since they were detected in only one epoch (2002 February 11).

The spatial distribution and proper motions of the masers around VLA~2 show a very compact symmetric  bipolar outflow in the northeast-southwest direction, consisting of 
two bow-shock-like structures separated by $\sim$0.4~arcsec ($\sim$360 au) moving in opposite directions, with VLA~2 located between them.
In Fig. 4 we show the spatio-kinematical distribution of the masers around VLA~2, with the radial velocities indicated in a colour scale (V$_\mathrm{LSR}$ $\simeq$ $-$18 to $-9$~km~s$^{-1}$), and the proper motion vectors represented by arrows. 
The magnitude
of the proper motions is in the range of $\sim$5 to 30~km~s$^{-1}$, although most of them have values $\gtrsim$ 20~km~s$^{-1}$ (see Table 1 and Fig. 4). 
Font, Mitchell \& Sandell (2001) report an ambient molecular cloud velocity V$_\mathrm{LSR}$ $\simeq$ $-$10.3~km~s$^{-1}$, obtained from $^{13}$CO and C$^{18}$O with a beam size of $\sim$14~arcsec. This velocity is similar to the mean radial velocity of the masers associated with VLA~2 (V$_\mathrm{LSR}$ $\simeq$ $-$11.6~km~s$^{-1}$; Table 1). Considering this, and that our data do not show any significant difference between the radial velocity of the northeastern and southwestern structures
(although Trinidad et al. 2004 reported a small segregation of $\sim$3~km~s$^{-1}$ in the radial velocity, with the northeastern group mainly redshifted and the southwestern one mainly blueshifted with respect to the ambient cloud velocity), we think that this is consistent with the masers to be located in the leading edge of projected bow shocks (Raga et al. 1997; Trinidad et al. 2013; see Appendix A).

In order to extend the study of the evolution of the H$_2$O maser bipolar outflow in VLA~2 up to a time-span of $\sim$3.6~yr, 
we have aligned the H$_2$O maser positions obtained with the VLBA by Marvel (2005)
on the single-epoch 2003 February 23, as well as those obtained with the VLA by Trinidad et al. (2004) on the single-epoch 1999 June 29, within the same reference frame of our aligned set of three VLBA epochs (2001 December 2, 2002 February 11, 2002 March 5) described in Section 2. For this alignment, we have first identified an H$_2$O maser feature clearly persisting in these five epochs. We identify the VLA feature number 4 listed by Trinidad et al. (2004) in their table 2 (V$_\mathrm{LSR}$ = $-$17.8~km~s$^{-1}$), with the two close VLBA features numbers 3 and 4 listed in our Table 1 (V$_\mathrm{LSR}$ = $-$17.5 and $-$18.2~km~s$^{-1}$, respectively), and the two close VLBA features 25 and 26 listed by Marvel (2005) in his table 4 (V$_\mathrm{LSR}$ = $-$17.8 and $-$17.7~km~s$^{-1}$, respectively). Then, we corrected the mean position of the two features 25 and 26 listed by Marvel (2005), assigning it to the location expected for epoch 2003 February 23 in our reference frame, obtained from the extrapolation of the mean position and proper motions of the features 3 and 4 of our data for epoch 2001 December 2 (Table 1), assuming they have moved with constant velocity during this time-span. This inferred spatial correction was applied to all the H$_2$O maser features of epoch 2003 February 23 detected by Marvel (2005).
We also applied a similar correction method to the position of the VLA feature number 4 (and therefore to all the data of epoch 1999 June 29; Trinidad et al. 2004) 
assigning the position expected from the extrapolation of our VLBA data. This must be taken as a rough alignment, mainly because the VLA observations are made with a beam size of 80~mas, therefore
large enough to contain several of the individual features observed with the VLBA, together with the assumption of constant velocity of the maser features during the long time-span of $\sim$3.6~yr.
In Fig. 5 we show the positions of the H$_2$O masers associated with VLA~2 measured with the VLA (epoch 1999 June 29) and VLBA (epochs 2002 February 11 and 2003 February 23). 
From this figure, wee see that, overall, the whole structure of the H$_2$O masers as observed in epoch 1999 June 29 is contained within the structure observed
in 2002 February 11, which in its turn, is contained within that observed in 2003 February 23, indicating that the structure has been continuously expanding during $\sim$3.6~yr. Measuring the angular separation between the northeastern and southwestern edges in 1999 June 29 ($\sim$373~mas) and in 2003 February 23 ($\sim$405~mas), we obtain an average expanding proper motion from a common center of $\sim$4.4~mas~yr$^{-1}$ ($\sim$19~km~s$^{-1}$). This
estimate is roughly consistent with the proper motions we have measured for the individual H$_2$O maser features in the shorter time-span of $\sim$0.25~yr (Table 1 and Fig. 4).

Finally, with respect to the proper motions of the H$_2$O maser features associated with VLA~3A, VLA~3B, or VLA~3 (A/B) (Table 1), we find that they do not show a preferential motion pattern in the sky from which we can extract any helpful physical information.
Therefore, in the next section we will focus only on discussing the results we have obtained in VLA~2.

\section{Discussion}

Our VLBA observations show a very compact, bipolar H$_2$O maser outflow centred on VLA~2, with the major axis of the maser distribution at p.a. 
$\sim$247$^{\circ}$ (see Fig. 4). Furthermore, the chain of masers to the southwest of VLA~2 can be fitted by a parabola. Using the offset positions of the H$_2$O maser features 1-20 in Table 1 that are shaping the southwestern maser structure of VLA~2, one obtains a parabola which represents the data with a standard deviation of $\sim$15~mas. However, if we exclude features 12 and 13, the deviation is reduced significantly to $\sim$4~mas. On the other hand, the length of the chain of masers is $\sim$200~mas, so the average distance between the masers and the parabolic fit represents only 2\% of its length. The p.a. of the axis of symmetry of the parabola is $\sim$230$^\circ$, similar to the p.a. of the major axis of the full maser distribution around VLA~2.  Another interesting result is that an extrapolation of the symmetry axis of the parabola passes at only $\sim$5~mas from VLA~2. If one forces the axis of the parabola to point exactly to this source, one obtains the following values for the best-fit parameters: coordinates of the focus, $x_0$ (RA offset) $ = -119$~mas, $y_0$ (DEC offset) $ = -100$~mas ($\pm3$~mas); distance between the focus and the tip of the parabola $h = 22 \pm 1$~mas; and p.a. of the symmetry axis  $= 230 \pm 3^\circ$ (see also Appendix A).

VLA~2 (NW2) is the weakest radio continuum source ($\sim$0.1~mJy) detected by Trinidad et al. (2004) at cm wavelengths in the region. It is not seen in the J (1.25~$\mu$m), H (1.635~$\mu$m), and K (2.2~$\mu$m) bands, but it is seen in the L' (3.77~$\mu$m) and M' (4.68~$\mu$m) bands, and it appears bright in the mid-infrared (11.8 and 17.65~$\mu$m) bands, as reported by Kato et al. (2011), who conclude that this is an embedded YSO. This source is fainter in the near- and mid-infrared than the nearby source NW1, coinciding with VLA~3A+3B (see Fig. 1).  These facts led Kato et al. (2011) to conclude that VLA~2 (NW2) is more embedded than NW1, or alternatively it has a lower luminosity.  From the spectral energy distribution of NW1, Kato et al. (2011) estimate a luminosity of $\sim$700~L$_{\odot}$ for this object. These authors  propose that NW1 is a YSO with spectral type B6-7, although this must be taken with caution, as it has been inferred assuming that it is a single main-sequence star, while NW1 appears to be an embedded binary since it 
has two components seen in radio continuum (VLA~3A and 3B), that are separated by $\sim$0.3~arcsec ($\sim$270 au; Trinidad et al. 2004). 
Kato et al. (2011) did not determine the spectral type of VLA~2 (NW2).
Here, using the VLA centimetre results, we discuss on the mass of VLA~2
under different assumptions. This source has a 3.6 cm flux density
$S_{3.6~cm} \simeq 0.1$ mJy (Trinidad et al. 2004). Assuming this emission
arises from an optically thin photoionised HII region, at a distance of
$d=0.9$ kpc, and with an electron temperature of 10$^4$~K, the required rate
of ionising photons is $\sim$$7\times 10^{42}~s^{-1}$ (e.g., Rodr\'\i guez et al. 1980).
This
ionising photon rate can be provided by a B4 ZAMS star (Thompson 1984)
with a luminosity of $\sim$700$L_\odot$, that would correspond to a main-sequence
star of $\sim$6 $M_\odot$ (Hohle, Neuh{\"a}user \& Schutz 2010).
These estimates are
likely upper limits given that part of the ionising photons could be
provided from shocks produced in the interaction of the outflow with
circumstellar material (Torrelles et al. 1985; Curiel et al. 1989).
Therefore, if we assume that the observed centimetre emission of VLA 2
traces a thermal radio jet, then, ionisation is expected to be produced by
shocks (Anglada 1995, 1996). From the correlation between the observed
$S_\nu d^2$ and bolometric luminosity of radio jet sources (see Fig. 5 in
Anglada 1995), we infer a bolometric luminosity of $\sim$$50~L_\odot$ for
VLA~2, that can be provided by an embedded protostar of a few solar
masses (e.g., Wuchterl et al. 2003; White et al. 2007). High angular
resolution observations with the recently improved sensitivity of the VLA
can determine whether VLA~2 shows the elongation and characteristics
expected for a thermal radio jet.

From the spatial extension of the bipolar H$_2$O maser outflow, $\sim$ 0.2~arcsec ($\sim$180 au; each side of the bipolar outflow, measured from VLA 2), and the expanding velocity of $\sim$20~km~s$^{-1}$ measured through the radial and proper motions of the masers, we estimate a kinematic age of $\sim$40~yr for the outflow. 
Given how unlikely it would be to observe this short-lived outflow if it happened only once in the lifetime of the YSO VLA~2, we propose that the 
observed outflow does not represent a single, steady process, but rather it is part of a repetitive pulsed jet event.
In this sense, Trinidad et al. (2004) proposed that VLA~2 is the powering source of the blueshifted [SII] jet observed in the region and extending to the south-west with p.a. $\sim$252$^{\circ}$ (Ray et al. 1990), because this position angle is similar to that of the major axis of the maser distribution measured with the VLA (p.a. $\sim$ 247$^{\circ}$; Fig. 1). Our VLBA H$_2$O maser observations give strong support to this interpretation because in addition to the spatial distribution of the masers, the bipolar motions are along a similar direction (p.a. $\sim$247$^{\circ}$; Fig. 4) as the [SII] jet.
Moreover, Fuente et al. (2001) propose that the 
redshifted counterpart of the blueshifted [SII] jet is 
the $^{12}$CO (J=1--0) outflow observed towards the north-east, extending up 
to $\sim$100~arcsec with a kinematic age of $\gtrsim$ 8$\times$10$^3$~yr. 
It might be possible that this extended outflow with higher kinematic age could be produced through repetitive short-lived ejections from VLA~2, as the one we find at small scales through our VLBA H$_2$O maser observations.

In fact, under the scenario of a wind-driven H$_2$O maser outflow in VLA~2, some important parameters, such 
as  the mass loss rate ($\dot M_w$) and 
terminal velocity ($V_w$), of the wind in these short-lived events can be estimated. The 
thermal
pressure inside the bow-shock-like maser structures is given by 
\begin{equation}
P_s = nkT, 
\end{equation}
where $n$ and $T$ are the particle density 
and temperature of the masing region, respectively. Setting this pressure of eq. (1) equal to that of the wind ejected within a solid angle $\Omega$,
\begin{equation}
P_s = \left(\frac{\dot M_w V_w}{\Omega R^2}\right),
\end{equation}
at a distance $R$ = 180 au (0.2~arcsec) from VLA~2, we then obtain:

$$\left(\frac{\dot M_w}{\rm M_\odot~yr^{-1}}\right)
\left(\frac{V_w}{\rm km~s^{-1}}\right) \simeq$$
\begin{equation}
1\times 10^{-4}
\left(\frac{n}{\rm 10^8~cm^{-3}}\right)\left(\frac{T}{\rm 500~K}\right)\left(\frac{R}{\rm 180~au}\right)^2\left(\frac{\Omega}{4\pi}\right).
\end{equation}
 Given that the
H$_2$O maser emission requires physical conditions of warm ($\sim$500
K) 
and dense ($\sim 10^8-10^9$~cm$^{-3}$)
molecular gas (e.g., Elitzur, Hollenbach \& McKee 1992; Claussen 2002; Hollenbach et al. 2013), we estimate from eq. (3)
\begin{equation}
\dot M_w V_w \simeq {\rm 10^{-4}-10^{-3} ~M_{\odot}~yr^{-1}~km~s^{-1}}\left(\frac{\Omega}{4\pi}\right).
\end{equation}
 From the geometry observed in the bipolar H$_2$O maser outflow, we estimate a solid angle of the wind $\Omega$ $\gtrsim$  0.28~sr, resulting from eq. (4) in a momentum rate in the range of $\sim$$10^{-3}-10^{-6}$ M$_{\odot}$~yr$^{-1}$~km~s$^{-1}$.

In addition, since the H$_2$O masers are shock-excited (e.g., Goddi et al. 2006; Surcis et al. 2011; Hollenbach, Elitzur \& McKee 2013),  we can also obtain the gas density outside the masing region (pre-shock density $n_0$), assuming a strong shock, as 
\begin{equation}
n_0 = n\left(\frac{a}{V_f}\right)^2,
\end{equation}
where $a = 1.4$~km~s$^{-1}$ is the sound speed inside the masing zone
(corresponding to $T = 500$~K), and $V_f$ is the outflow velocity traced by the masers ($\sim$20~km~s$^{-1}$), resulting in
\begin{equation}
 \left(\frac{n_0}{\rm cm^{-3}}\right) \simeq 5\times10^5\left(\frac{n}{\rm 10^8~cm^{-3}}\right). 
 \end{equation}
 We think that this is a reasonable pre-shock density value for the molecular core of the star-forming LkH$\alpha$~234 region where the different YSOs are embedded.

As mentioned in Section 1, short-lived episodic ejection events have been found towards massive YSOs through multi-epoch VLBI H$_2$O maser observations.  With our VLBA observations we have extended that kind of study to intermediate-mass YSOs, providing an example of a short-lived episodic ejection event associated with the source VLA~2 (NW2) in the star-forming LkH$\alpha$~234 region. This could indicate that the accretion process in this YSO is also a non-steady process, but undergoing episodic increases in the accretion rate from the expected associated circumstellar disk, as it has been observed in low-mass YSOs through FU Orionis events or episodic ejection in the jets associated with HH objects (e.g., Hartmann \& Kenyon 1996; Schwartz \& Greene 2003; Lee et al. 2007; Greene, Aspin \& Reipurth 2008; Pech et al. 2010; Stamatellos, Whitworth \&  Hubber 2012). Actually, episodic accretion has been proposed to solve the so-called "luminosity problem" in the modelling of Class I sources (Kenyon et al. 1990; Osorio et al. 2003).

Finally, our data, showing the presence of a compact ($\sim$360~au) well collimated bipolar outflow in VLA~2 (NW2), predict the presence of an accretion disk at these scales. 
In this sense, the central masers which are distributed along a linear structure with a size of $\sim$10~mas, first reported by Marvel (2005) and being more evident in the VLBA epoch of 2003 February 23 (see Fig. 5), could trace the innermost 
interaction region of the jet with its disk as suggested by this author.
The presence of the disk can be tested through (sub)mm dust continuum and thermal line observations.
These observations, together with sensitive radio continuum observations, would allow us to characterise not only the nature of the VLA~2 (NW2) object, but also the nature of the cluster of YSOs found in this intermediate-mass star-forming region.

\section{Conclusions}

We report VLBI H$_2$O maser observations towards the intermediate-mass star-forming LkH$\alpha$~234 region.
We have detected H$_2$O maser emission towards four YSOs (VLA~1, VLA~2, VLA~3A, and VLA~3B) within a region of $\sim$4~arcsec ($\sim$3,600~au) in size, 
indicating outflow activity from these objects. In particular, the spatio-kinematical distribution of the masers reveal a remarkable very young, compact  bipolar
H$_2$O maser outflow associated with the intermediate-mass YSO VLA~2, which is located at the centre of the outflow. 
This outflow is formed by two bow-shock-like structures moving in opposite directions. 
The size of the outflow (0.2 arcsec = 180~au; measured from VLA~2), together with the expanding velocity of the masers with respect to the central source ($\sim$20~km~s$^{-1}$), gives a kinematic age of $\sim$40~yr. We interpret this outflow as driven by a short-lived episodic jet ejection from VLA~2, rather than a steady outflow. Short-lived episodic ejections have been also observed in massive YSOs through VLBI H$_2$O maser observations. With the results presented in this paper we show now that this behaviour may also occur in intermediate-mass YSOs, probably due to episodic increases in the accretion rate as observed and firmly established in low-mass YSOs. In this way, our observations predict the presence of an accretion disk around VLA~2 forming a disk-YSO-jet system similar to what is observed in low- and high-mass YSOs. This might be studied with sensitive radio continuum and (sub)mm dust continuum and spectral line observations at scales of a few tenths of an arcsec.

Our results support VLA~2 as the powering source of the extended outflow ($\sim$100~arcsec = 9,000~au) seen in [SII]/CO. It is possible that this extended outflow, with kinematic age $\gtrsim$ 8$\times$10$^3$~yr, might be driven by short-lived episodic ejections from VLA~2 as the very recent one traced by the VLBI H$_2$O maser observations.

\section*{Acknowledgments}

We are very grateful to Eri Kato for providing us the near- and mid-infrared images shown in this paper (Fig.1). 
We thank the referee, Kevin Marvel, for his valuable comments and suggestions on the manuscript.
GA, RE, JFG, JMG, and JMT acknowledge support from MICINN (Spain) AYA2011-30228-C03 grant (co-funded with FEDER funds). JC acknowledges support from CONACyT grant 61547. 
SC acknowledges the support of DGAPA, UNAM, CONACyT (M\'exico), and CSIC (Spain).
CC-G and LFR acknowledges the support of DGAPA, UNAM, and CONACyT (M\'exico).
MAT acknowledges support from CONACyT grant 82543. JMG, RE, and JMT acknowledge support from AGAUR (Catalonia) 2009SGR1172 grant. The ICC (UB) is a CSIC-Associated Unit through the ICE (CSIC).

\begin{appendix}

\section{Bow shock approach}

As shown in Section 3, the chain of masers to the southwest of VLA~2 can be fitted by a parabola with high accuracy (excluding features 12 and 13 of Table 1), with coordinates of the focus 
$x_0$ (RA offset) $ = -119$~mas, $y_0$ (DEC offset) $ = -100$~mas ($\pm3$~mas); distance between the focus and the tip of the parabola $h = 22 \pm 1$~mas; and position angle of the symmetry axis of the parabola $p.a. = 230^\circ \pm 3^\circ$.
The proper motions of the masers can also be roughly modelled if we assume a moving bow shock of paraboloidal shape (the projection of the bow shock on the plane of the sky is also a parabola). The fact that the mean radial velocity of the masers is similar to the velocity of the ambient molecular cloud (Section 2.1.2) would agree with the scenario of the chain of maser to be located in the leading edge of the projected bow shock (Raga et al. 1997). From Raga et al. (1997) and Trinidad et al. (2013) it is easy to show that, locally, the post-shock gas moves in the direction normal to the shock. The projection of this velocity on the plane of the sky for points along the observed edge of the bow shock, in a ($x'$, $y'$) reference frame centred on the focus of the parabola and aligned with the projected bow-shock shape (with the $x'$-axis parallel to the symmetry axis, and $y'$-axis perpendicular to it) is,

\begin{equation}
v = \frac{2h}{(4h^2 + y'^2)^{1/2}} v_{bs}\cos \phi
\end{equation}

\noindent where $v_{bs}$ is the velocity of the bow shock, $\phi$ the angle between its direction of
motion and the plane of the sky, and $y'$ the distance between the maser and the symmetry axis of the parabola.
From Trinidad et al. (2013), we can also see that the components of this velocity along ($x'$) and perpendicular ($y'$) to the symmetry axis are given by

\begin{equation}
v_{x'} = v^0_{x'} + \frac{4h^2}{4h^2 + y'^2} v_{bs}\cos\phi
\end{equation}

\begin{equation}
v_{y'} = v^0_{y'} + \frac{2hy'}{4h^2 + y'^2} v_{bs}\cos\phi
\end{equation}

\noindent where $v^0_{x'}$ and $v^0_{y'}$ represent the components of a possible motion (with respect to
the observer) of the environment within which the bow shock is moving. The total velocity on the plane of the sky with respect to the observer is,

\begin{equation}
v_t = \sqrt{v^2_{x'} + v^2_{y'}}
\end{equation}

Using the transformation equations,

\begin{eqnarray}
\begin{tabular}{ll}
$  y'  $ & = $ -(\Delta x - x_0)\sin \theta_0 + (\Delta y - y_0)\cos \theta_0 $ \\
$ v_{x'} $  & = $ v_x \cos \theta_0 + v_y \sin \theta_0 $ \\
$ v_{y'} $ &  = $ -v_x \sin \theta_0 + v_y \cos \theta_0 $
\end{tabular}
\end{eqnarray}

\noindent we obtain the distance $y'$ and the $v_{x'}$, and $v_{y'}$ velocities of the masers from the
observed proper motions ($v_{x}$, $v_{y}$) and sky offset positions ($\Delta x$, $\Delta y$) given in Table 1. In the equations above, the angle $\theta_0 = \pi/2 - p.a.$,  where $p.a.$, $x_0$, and $y_0$ are those derived from the fit to the maser positions with a parabola with symmetry axis pointing to VLA~2. Fig. A1 shows the resulting velocities as a function of the distance $y'$ and the fit of equations (A2)-(A4) to these data (with $h = 22$ mas). Table A1  gives the values of the best-fit parameters. 

The dispersion of the values of the fitting parameters that we obtain indicates that a ``simple'' bow shock model with a paraboloidal shape, although it can be considered as a rough representation of the observations, cannot reproduce exactly all the observed spatio-kinematical properties of the H$_2$O masers in the region.

\begin{table}
\caption{Parameters of the parabolic fit to the bow shape}
\begin{center}
\begin{tabular}{cccc}
\hline\hline
$v^0_{x'}$ & $v^0_{y'}$ & $v_{bs}\cos \phi$ \\
(km~s$^{-1}$) & (km~s$^{-1}$) & (km~s$^{-1}$)\\
\hline
2 & ... & 17 & from eq. (A2) \\
... & $-$3 & 10 & from eq. (A3) \\
0.4 & $-$10 & 17 & from eq. (A4) \\
\hline
\end{tabular}
\end{center}
\end{table}

\begin{figure}
\centering
\begin{tabular}{c}
\includegraphics[width=82mm, clip=true]{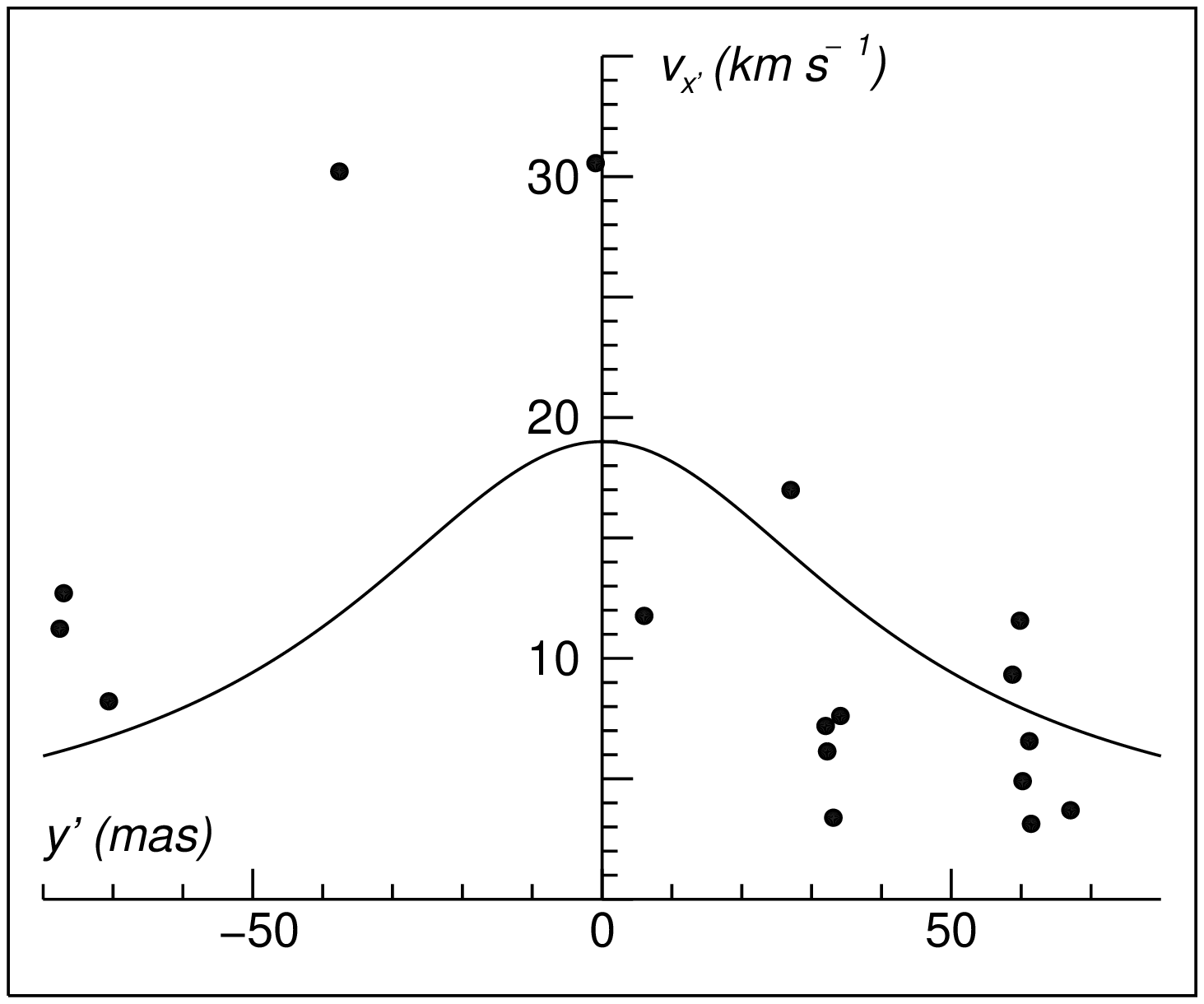}\\
\includegraphics[width=82mm, clip=true]{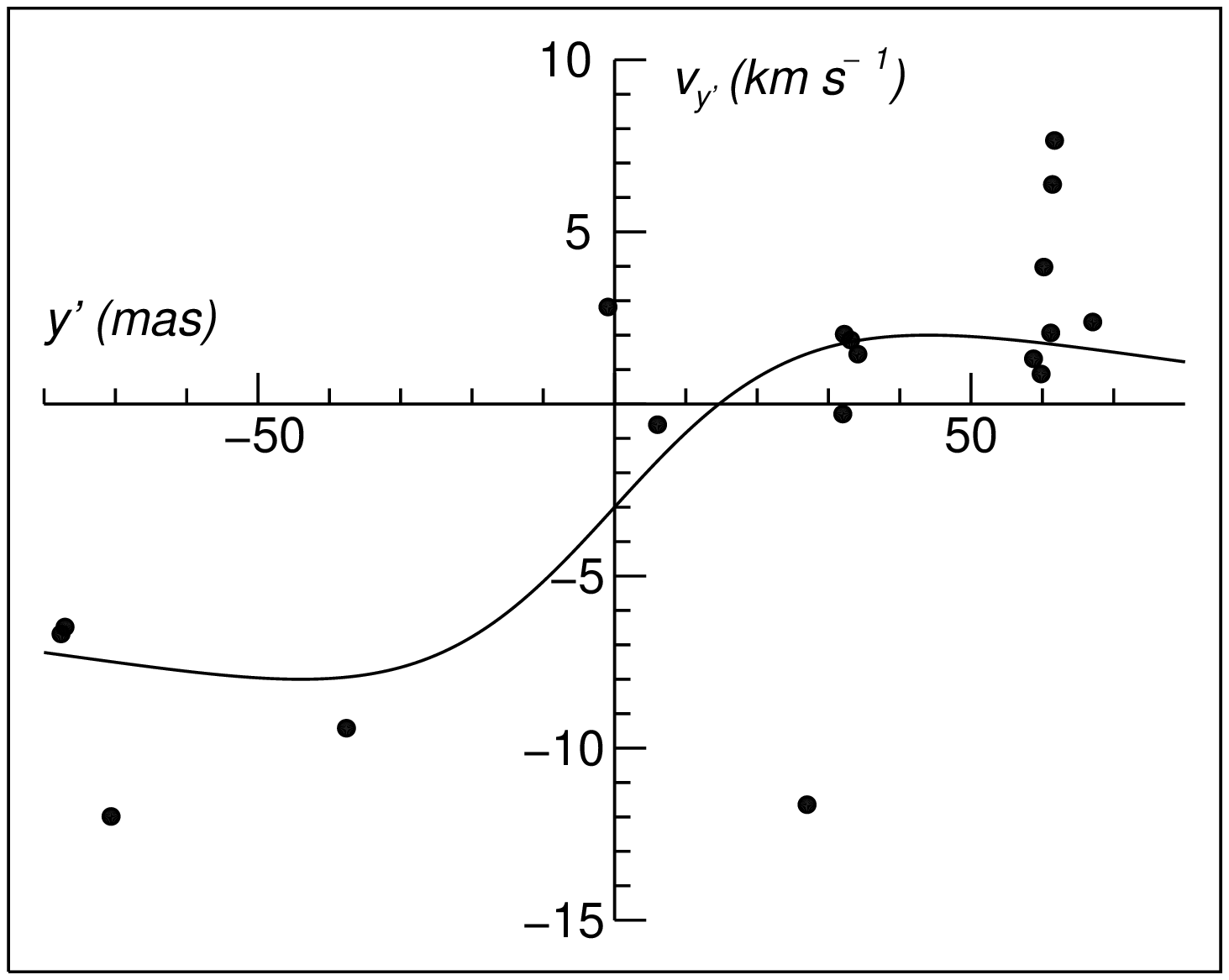}\\
\includegraphics[width=82mm, clip=true]{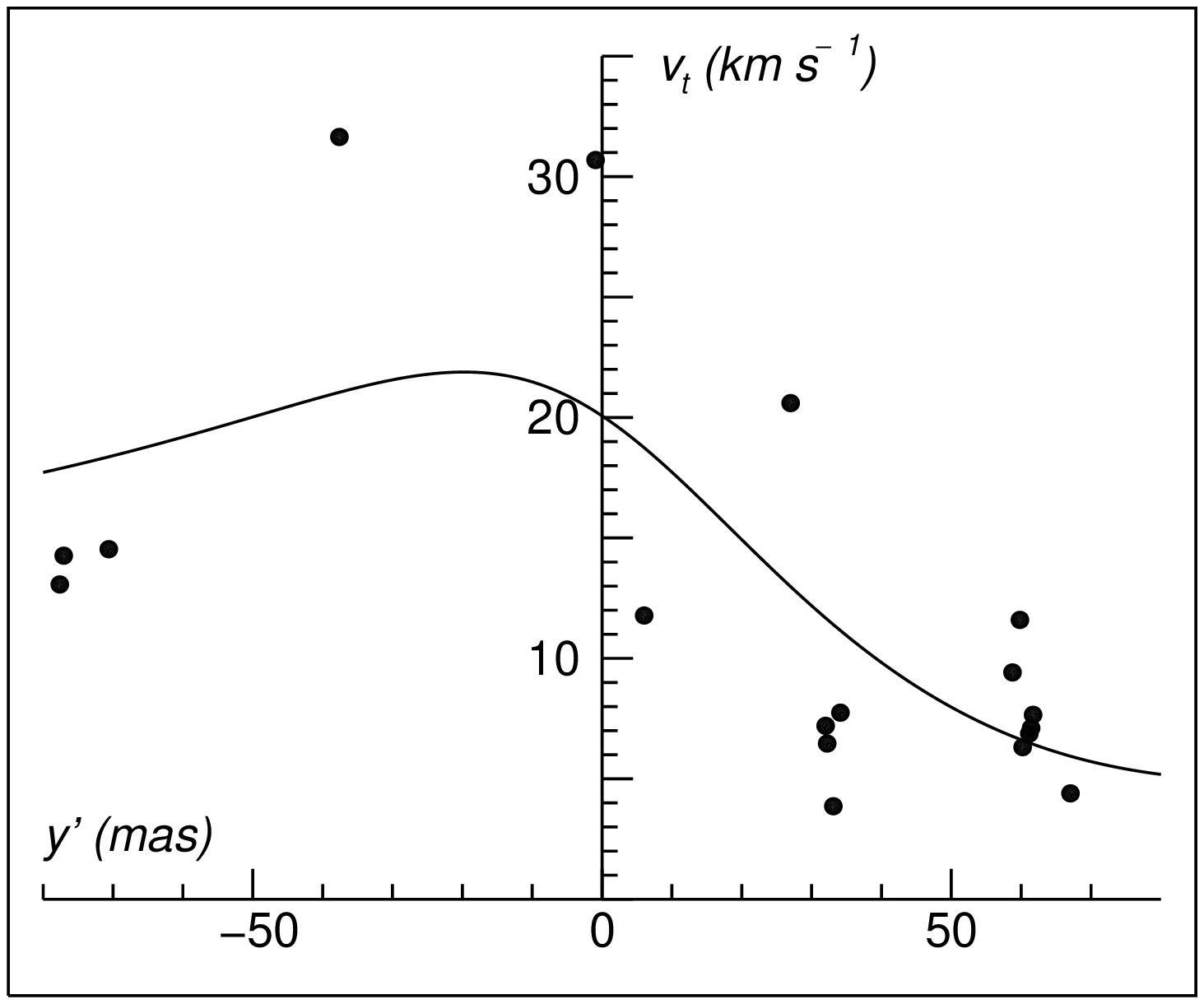}
\end{tabular}
\caption{ The dots show the proper motions $v_{x'}$, $v_{y'}$, and
$v_t$ of the individual H$_2$O maser features 
as a function of the $y'$ coordinate (measured perpendicular to
the symmetry axis of the projected bow shock). The
solid lines correspond to the least squares fits described
in the text.} 
\end{figure}

\end{appendix}

\label{lastpage}

\bsp

\end{document}